# A New Approach to Calculate the Transport Matrix in RF Cavities[*]

Yu. Eidelman[#], BINP, Novosibirsk 630090, Russia, N. Mokhov, S. Nagaitsev, N. Solyak, FNAL, Batavia, IL 60510, U.S.A

*Abstract*

A realistic approach to calculate the transport matrix in RF cavities is developed. It is based on joint solution of equations of longitudinal and transverse motion of a charged particle in an electromagnetic field of the linac. This field is a given by distribution (measured or calculated) of the component of the longitudinal electric field on the axis of the linac. New approach is compared with other matrix methods to solve the same problem. The comparison with code ASTRA has been carried out. Complete agreement for tracking results for a TESLA-type cavity is achieved. A corresponding algorithm will be implemented into the MARS15 code.

## TRANSVERSE MOTION OF THE CHARGE PARTICLE IN THE RF CAVITY

*Equation*

Let us consider a cavity with the given distribution of the longitudinal field along its axis, so that the function $E_z(z)$ is known. It means that the radial electrical field $E_r(z)$ and azimuthal magnetic one $H_\varphi(z)$ will affects an electron which is moving in the cavity according to the following Maxwell's equations:

$$\vec{E}_r(z) = -\frac{r}{2}\frac{\partial E_z(z)}{\partial z}\vec{e}_r \text{ and } \vec{B}_\varphi(z) = \frac{r}{2c}\frac{\partial E_z(z)}{\partial t}\vec{e}_\varphi = \frac{\beta r}{2}\frac{\partial E_z(z)}{\partial z}\vec{e}_\varphi,$$

where $r$ is the distance from the cavity axis and $\beta$ is the relative velocity of the electron. These transverse fields result in a Lorenz force which is radially affecting the electron (with mass $m$ and absolute charge $q$) and equals

$$\vec{F}_r = -q(E_r - \beta B_\varphi)\vec{e}_r = \frac{q(1+\beta^2)}{2}r\frac{\partial E_z}{\partial z}\vec{e}_r. \quad (1)$$

In the case of an axially-symmetric cavity it is possible to consider the electron motion in the plane $(x,z)$ and after simple transformation of the expression (1) one can obtain the following equation for transverse motion of the particle:

$$x'' + \frac{1}{\beta^2}\frac{\gamma'}{\gamma}x' = \frac{q}{mc^2}\frac{1+\beta^2}{2\beta^2\gamma}x\frac{\partial E_z}{\partial z}. \quad (2)$$

Here the prime means derivative with respect to z. Usually, the motion of electron is considered as ultra relativistic, so that $(1+\beta^2)/2\beta^2 \approx 1$, but we will not restrict ourselves to this case only.

*Standard Matrix Consideration [1, 2]*

The standard approach is as follows. The electromagnetic field of RF cavity includes a few higher spatial (temporal) harmonics. For this reason it is possible to present the motion of the charge particle as sum of two components: smooth ("slow") and "fast" and apply the matrix approach to solve the equation (2). After averaging over time, significantly exceeds the characteristic time of the fast component, and after the necessary transformations, it is possible to reduce the equation (2) to the following form:

$$x''(z) + \frac{1}{\beta^2}\frac{\gamma'}{\gamma}x'(z) + \frac{1}{2}\left(\frac{qE_0}{mc^2}\frac{1+\beta^2}{4\beta^2\gamma}\right)^2 x(z)\eta(\Delta\phi) = 0. \quad (3)$$

Some special factor $\eta(\Delta\phi)$ appears in this equation to takes into account an RF-field structure in the cavity and a phase shift $\Delta\phi$ of the particle when it enters to the cavity.

Solution of equation (3) for ultra relativistic particle in the case of the "pure" (without other spatial harmonics) $\pi-$ mode of the field in the cavity can be written using the so-called matrix Chambers [3, 4] (which already takes into account the effect of the edge of the transverse focusing [1, 5] for the entrance/exit of the cavity):

$$\begin{pmatrix} x \\ x' \end{pmatrix}_f = M^{Ch} \begin{pmatrix} x \\ x' \end{pmatrix}_i \quad (4)$$

with ($\alpha = \frac{\sqrt{\eta(\Delta\phi)/8}}{\cos(\Delta\phi)}\ln\frac{\gamma_f}{\gamma_i}$). The matrix $M^{Ch}$ equals

$$\begin{aligned}
M^{Ch}_{11} &= \cos\alpha - \sqrt{2}\cdot\cos(\Delta\phi)\cdot\sin\alpha, \\
M^{Ch}_{12} &= \cos\alpha - \sqrt{2}\cdot\cos(\Delta\phi)\cdot\sin\alpha, \\
M^{Ch}_{21} &= -\left(\frac{\gamma'}{\gamma_f}\right)\cdot\left[\frac{\cos(\Delta\phi)}{\sqrt{2}} + \frac{1}{\sqrt{8}\cos(\Delta\phi)}\right]\cdot\sin\alpha, \\
M^{Ch}_{22} &= \left(\frac{\gamma_i}{\gamma_f}\right)\cdot\left[\cos\alpha + \sqrt{2}\cdot\cos(\Delta\phi)\cdot\sin\alpha\right].
\end{aligned} \quad (5)$$

Some other matrix representations of particle motion in the cavity are discussed in details in the paper [6].

---

*Work supported by US DoE.
[#]eidelyur@inp.nsk.su

## NEW MATRIX APPROACH

The main feature of this approach is to use the known distribution of the RF cavity electric field $E_z(z,t)$, the paraxial character of particle motion and do not restrict the consideration to ultra relativistic energies of the particle only.

### Equations and solution

To do this, the following standard equation for the longitudinal motion (acceleration) of a particle in the field of the cavity must be used:

$$\frac{d\gamma}{dt} = \frac{q\beta}{mc} E_z(z,t), \quad (6)$$

as well as to convert equation (2) to a more convenient form [11]:

$$m\gamma\ddot{x} + \frac{q\beta}{c} E_z \dot{x} + \frac{q}{2}\left(\frac{\partial E_z}{\partial z} + \frac{\beta}{c}\frac{\partial E_z}{\partial t}\right) x = 0. \quad (7)$$

For RF cavity with the transverse TM-mode

$$E_z(z,t) = E_0(z)\cos(\omega t + \Delta\phi) \quad (8)$$

the equation (7) takes the form

$$m\gamma\ddot{x} + \frac{q\beta}{c} E_0 \cos(\omega t + \Delta\phi)\dot{x} + \frac{q}{2}\cdot$$

$$\left[\frac{\partial E_0}{\partial z}\cos(\omega t + \Delta\phi) - \frac{\omega\beta}{c} E_0 \sin(\omega t + \Delta\phi)\right] x = 0$$

and the equation (6) can be integrated, so that relative gain $\Delta\gamma/\gamma$ of the particle energy while moving through the cavity during the time interval $t \div t + \Delta t$ is equal to

$$\frac{\Delta\gamma}{\gamma} = \frac{q\bar{\beta}\bar{E}_0}{mc\bar{\gamma}} \cdot \Delta t \cdot \text{sinc}\frac{\omega\cdot\Delta t}{2}\cdot\cos[\omega\bar{t} + \Delta\phi]. \quad (9)$$

All "bar-values" in this expression are referred to the moment $t + \Delta t/2$.

It is convenient to use a "length" $\tau = ct$, wave number $k_0 = \omega/c$ and express the amplitude of the electric field $\tilde{E}_0 = qE_0/mc^2$. Then one can find the final equation for the transverse motion (a sign «'» means now the derivative over length $\tau$):

$$x'' + \frac{\beta}{\gamma}\tilde{E}_0\cos(k_0\tau+\Delta\phi)x' + \frac{1}{2\gamma}\cdot$$

$$\left[\frac{\partial \tilde{E}_0}{\partial z}\cos(k_0\tau+\Delta\phi) - k_0\beta\tilde{E}_0\sin(k_0\tau+\Delta\phi)\right]x = 0. \quad (10)$$

To integrate this equation over $\tau$ it is necessary to present the whole integration range as a number of subintervals (slices) $\tau \div \tau + \Delta\tau$ and corresponding slices $z \div z + \Delta z$. On each of these subintervals one can neglect a change of parameters $\beta$, $\gamma$, as well as of the values of the field $\tilde{E}_0(z)$ and its derivative. In this approach, instead of equation (10) one has the simplest equation of the second order with "constant" coefficients:

$$x'' + zx' + bx = 0 \quad (11)$$

where

$$a = \frac{\bar{\beta}}{\bar{\gamma}}\tilde{E}_0(\bar{z})\cos(k_0\bar{\tau}+\Delta\phi),$$

$$b = \frac{1}{2\bar{\gamma}}\left[\frac{\partial \tilde{E}_0(\bar{z})}{\partial z}\cos(k_0\bar{\tau}+\Delta\phi) - k_0\bar{\beta}\tilde{E}_0(\bar{z})\sin(k_0\bar{\tau}+\Delta\phi)\right]$$

with $\bar{\beta} = \beta(\bar{z})$, $\bar{\gamma} = \gamma(\bar{z})$ and $\bar{\tau}$, $\bar{z}$ are the centers of the slices. It is quite easy to find a solution to this equation for the coordinate $x_f$ and angle $x'_f$ at the exit of the cavity using their values $x_i$, $x'_i$ at the entrance:

$$x(\tau) = \frac{-\alpha_2 x_i + x'_i}{\alpha_1 - \alpha_2}e^{\alpha_1\tau} + \frac{\alpha_1 x_i - x'_i}{\alpha_1 - \alpha_2}e^{\alpha_2\tau},$$

$$x'(\tau) = \frac{-\alpha_2 x_i + x'_i}{\alpha_1 - \alpha_2}\alpha_1 e^{\alpha_1\tau} + \frac{\alpha_1 x_i - x'_i}{\alpha_1 - \alpha_2}\alpha_2 e^{\alpha_2\tau}$$

Where $\alpha_{1,2} = (-a \pm \sqrt{a^2 - 4b})/2$. These expressions allow one to find the desired matrix of transformation of the coordinate vector during particle passage through the cavity. Let us use the coefficient $\alpha_{1,2}$ and introduce the following parameters:

$$\varepsilon = \frac{1}{2\bar{\gamma}}\sqrt{\begin{array}{l}\bar{\beta}^2\tilde{E}_0^2(\bar{z})\cos^2(k_0\bar{\tau}+\phi) - \\ \left[\frac{\partial \tilde{E}_0(\bar{z})}{\partial z}\cos(k_0\bar{\tau}+\phi) - k_0\bar{\beta}\tilde{E}_0(\bar{z})\sin(k_0\bar{\tau}+\phi)\right]\end{array}}; \quad (12)$$

$$\delta = \frac{\bar{\beta}}{2\bar{\gamma}}\tilde{E}_0(\bar{z})\cos(k_0\bar{\tau}+\phi),$$

So, $\alpha_{1,2} = -\delta \pm \varepsilon$ and after simple manipulations the following result will be found for the matrix $M$ of the slice of the cavity with length $\Delta\tau$:

$$\left.\begin{array}{l}M_{12} = e^{-\delta\cdot\Delta\tau} \cdot \dfrac{sh(\varepsilon\cdot\Delta\tau)}{\varepsilon},\\[4pt] M_{11} = e^{-\delta\cdot\Delta\tau}\cdot ch(\delta\cdot\Delta\tau) + M_{12}\cdot\delta,\\[4pt] M_{21} = (\varepsilon^2 - \delta^2)\cdot M_{12},\\[4pt] M_{22} = M_{11} - 2\cdot M_{12}\cdot\delta.\end{array}\right\} \quad (13)$$

It is very simple to calculate the determinant of the transport matrix $M$ of the cavity: $\det M = e^{-2\delta\cdot\Delta\tau} = e^{-\Delta\gamma/\gamma}$. As mentioned above, during slicing of the whole cavity in order to integrate the transverse motion of the particle it is necessary to take into account that for each slice the relative acceleration rate must be small, i.e. for all subintervals with length $\Delta\tau$, the value $\Delta\gamma/\gamma \ll 1$, so it is possible to replace the direct integration by a solution which uses the matrix approach.

## *Verification of the new approach*

To verify this approach, the code MatLab Dark Current (MLDC) was created. This code realizes two possibilities: direct integration of equation (11) by method Runge-Kutta with a fixed time step (4$^{th}$ order; function ode45 from the MatLab package) and matrix approach (using expressions (13)) for this equation. To calculate the acceleration rate the expression (9) was used.

The results of simulations with the code MLDC were compared with the results (naturally, for the same data), received while using the code ASTRA (A Space charge TRacking Algorithm) [7].

To compare both codes, the TESLA-type cavity is used with field amplitude $E_0 = 36.815$ MV/m.

Results of the scanning over a phase are shown for both codes are shown in Fig. 1.

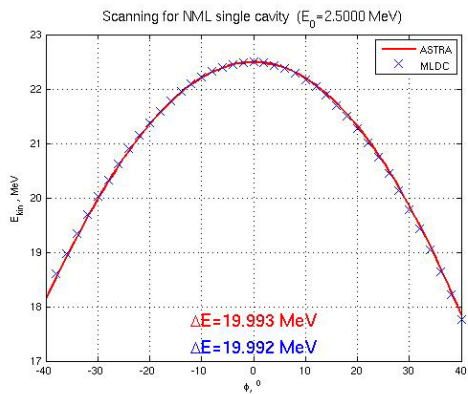

Fig. 1. Energy gain in the NML-cavity depending on phase of the particle.

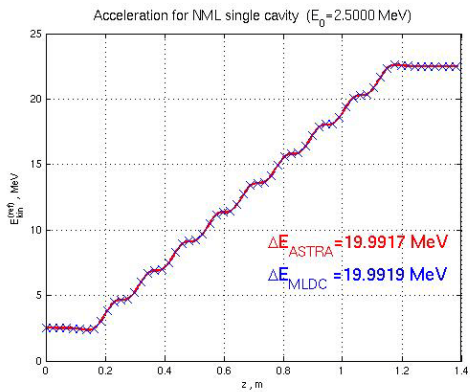

Fig. 2. Acceleration in the Tesla-type cavity.

The Fig. 2 is illustrated the process of the acceleration and gives the same results for both codes.

Next figures demonstrate the result of tracking with both codes in the cases of DI (direct integration) and MA (matrix approach).

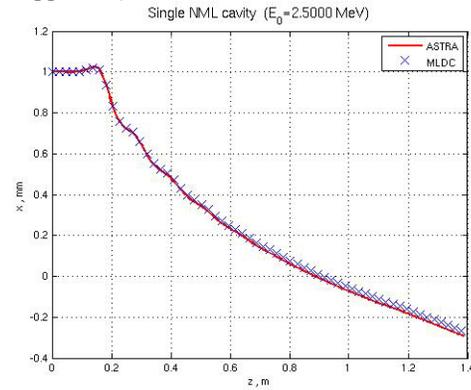

Fig. 3. Particle's track (DI approach). $E_0 = 2.5$ MeV.

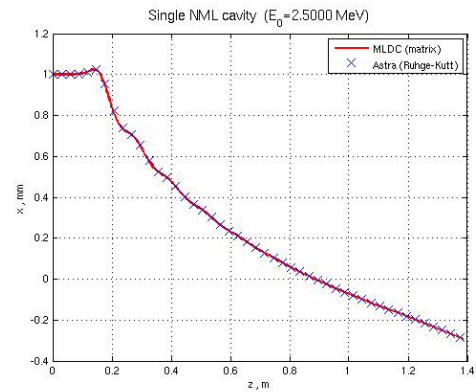

Fig. 4. Particle's track (MA method). $E_0 = 2.5$ MeV.

Complete agreement between all results is achieved. It proves the validity of the code MLDC and approaches used to create it.

## CONCLUSSIONS

A realistic approach to calculate the transport matrix in RF cavities is developed. Complete agreement for tracking results with existed code ASTRA is achieved. New algorithm will be implemented into MARS15 code.